# Bounding the Risk Difference Bias Given Incomplete Parameters for Externally Adjusting Uncontrolled Confounding

Onyebuchi A. Arah[1,2,3,4,5]

[1]Department of Epidemiology, Fielding School of Public Health, University of California, Los Angeles (UCLA), Los Angeles, USA

[2]Department of Statistics and Data Science, Division of Physical Sciences, University of California, Los Angeles (UCLA), Los Angeles, USA

[3]Practical Causal Inference Lab, University of California, Los Angeles (UCLA), Los Angeles, USA

[4]California Center for Population Research, UCLA, Los Angeles, USA

[5] Research Unit for Epidemiology, Department of Public Health, Aarhus University, Aarhus, Denmark

Correspondence: Dr. O. A. Arah, Department of Epidemiology, UCLA School of Public Health, Los Angeles, CA 90095-1772, USA

Email: arah@ucla.edu

ORCiD: 0000-0002-9067-1697

Running title: Bounding the Risk Difference Bias

**ABSTRACT**

Bias due to uncontrolled confounding can be analyzed using several external adjustment formulas. Like other bias expressions, the risk difference bias formula involving an uncontrolled polytomous confounder requires only three externally obtainable parameters: confounder prevalences among the exposed and unexposed and the confounder-outcome risk difference. This paper presents two-parameter bounds on the risk difference bias expression when only one of the three required bias parameters is unavailable. Each bound is sharp over the full range of the unknown parameter. Some numerical examples with graphs are provided.

## INTRODUCTION

In epidemiologic research, uncontrolled confounding remains an important but poorly appreciated problem (1-9). Although nonrandomized or observational studies are most susceptible to this problem, randomized trials can also suffer from uncontrolled confounding when randomization fails to render the comparison groups exchangeable, as seen with nonrandom noncompliance (10-13). Several methods have been developed to handle uncontrolled confounding. One method involves directly bounding the target causal effect using the potential outcomes framework, implemented via linear programming (11, 14). The other, more commonly used, method involves external adjustment formulas for simple or thorough sensitivity analysis (1-9, 15-18). These external adjustment or bias formulas typically require three parameters: the confounder-outcome effect measure, the confounder-exposure association, and the confounder prevalence among the unexposed. The confounder-exposure association is only a function of the confounder prevalences among the exposed and unexposed, and therefore, can be replaced by the confounder prevalences among the exposed to produce basic bias expressions (17).

In situations where only partial data or incomplete expert knowledge is available for one of the 3 basic parameters, it is still possible to estimate bounds on the bias factors (7, 19). In this paper, we develop bounds for the risk difference bias expression under uncontrolled confounding in cohort studies (17, 20). We first show that, when only one of the three parameters required for calculating the RD bias expression is unavailable, an expression containing the remaining parameters in the form most closely resembling the original bias expression provides the maximum or minimum bound for bias. Throughout this article, we assume that the unobserved confounder-adjusted effect would have been standardized to the exposed group (although extension to other standard populations is straightforward). It is also assumed that there are no other sources of bias. No homogeneity assumption is needed, as relevant expressions leave the exposure-confounder interaction unrestricted. Random error in the bounds, which could result from the data, is ignored, although this can be easily included in the analysis.

## BOUNDS ON THE RISK DIFFERENCE BIAS

In line with previous results (17, 20), it can be shown that when the exposed group is the standard population, the bias expression ($Bias_{RD}$) for the risk difference is given by

$$RD_{DX+} - RD_{DX,E} = \sum_{i=1}^{K} RD_{DZ0i}(P_{Z1i} - P_{Z0i}) \quad [1]$$

where

$D$ is a binary outcome (1 = outcome occurred, 0 not),

$X$ is a binary exposure (1 = exposed, 0 not),

$Z$ is an unmeasured polytomous confounder with K categories: $i = 1, \ldots, K$

$RD_{DX+}$ is the crude risk difference and is given by $\Pr(D = 1|X = 1) - \Pr(D = 1|X = 0)$,

$RD_{DX,E}$ is the unobserved confounder-adjusted risk difference standardized to the exposed group and is given by $\Pr(Z = i|X = 1)\cdot[\Pr(D = 1|X = 1, Z = i) - \Pr(D = 1|X = 0, Z = i)]$,

$RD_{DZ0i}$ is the confounder-outcome risk difference among the unexposed and is estimated as $\Pr(D = 1|Z = i, X = 0) - \Pr(D = 1|Z = 1, X = 0)$; being a difference of two probabilities, it necessarily lies in $[-1, 1]$, a restriction on which all of the bounds below depend,

$P_{Z1i}$, the prevalence of each confounder category among the exposed, is given by $\Pr(Z = i|X = 1)$, and

$P_{Z0i}$, the prevalence of each confounder category among the unexposed, is given by Pr(Z = $i$|X = 0).

We will now vary each of the three parameters, $RD_{DZ0i}$, $P_{Z1i}$, and $P_{Z0i}$, one at a time, to estimate the maximum upper and minimum lower bounds for $Bias_{RD}$. These bounds will be 'maximum' or 'minimum' because, given the two known parameters, $Bias_{RD}$ will never be larger or smaller than the implied limits, regardless of the value taken by the unknown parameter.

**Unknown confounder-outcome risk difference (*$RD_{DZ0i}$*):** When the exposure-specific confounder-outcome risk difference ($RD_{DZ0i}$) is the unobtainable parameter, and it is assumed to be greater than null (that is, $RD_{DZ0i} \geq 0$), then $Bias_{RD}$ can be shown to be bounded as follows (see Appendix):

$$0 \leq Bias_{RD} \leq \sum_{i=2}^{K}(P_{Z1i} - P_{Z0i}) \qquad \text{if } P_{Z1i} \geq P_{Z0i} \qquad [2]$$

$$\sum_{i=2}^{K}(P_{Z1i} - P_{Z0i}) \leq Bias_{RD} \leq 0 \qquad \text{if } P_{Z1i} \leq P_{Z0i} \qquad [3]$$

where $i$ = 2, …, $K$ refers to the set of all $Z$-strata except the reference category $i$ = 1, where $RD_{DZ0i} = 0$. Throughout, a condition of the form $P_{Z1i} \geq P_{Z0i}$ (or $P_{Z1i} \leq P_{Z0i}$) is to be read as applying to $i$ = 2, …, $K$ only. Inequality [2] gives the bounds on the nonnegative $Bias_{RD}$, while inequality [3] gives the bounds on the nonpositive bias. Now, assume that $RD_{DZ0i} \leq 0$, then the bounds for $Bias_{RD}$ are given by

$$\sum_{i=2}^{K}(P_{Z0i} - P_{Z1i}) \leq Bias_{RD} \leq 0 \qquad \text{if } P_{Z1i} \geq P_{Z0i} \qquad [4]$$

and

$$0 \leq Bias_{RD} \leq \sum_{i=2}^{K}(P_{Z0i} - P_{Z1i}) \qquad \text{if } P_{Z1i} \leq P_{Z0i}. \qquad [5]$$

Combining the inequalities [2] and [4] yields the minimum and maximum bounds on $Bias_{RD}$ given only $P_{Z1i} \geq P_{Z0i}$:

$$\sum_{i=2}^{K}(P_{Z0i} - P_{Z1i}) \leq Bias_{RD} \leq \sum_{i=2}^{K}(P_{Z1i} - P_{Z0i}). \quad [6]$$

Figure 1 presents a graphical illustration of this bound for a three-category confounder where $P_{Z13} = 0.50$, $P_{Z12} = 0.40$, $P_{Z11} = 0.10$, $P_{Z03} = 0.40$, $P_{Z02} = 0.35$, $P_{Z01} = 0.25$, and both $RD_{DZ03}$ and $RD_{DZ02}$ are allowed to vary. The absolute values of both the lower and upper bounds in inequality 6 are equal since, over the full range of $RD_{DZ0i}$, $Bias_{RD}$ is symmetrical about its null, zero. The relationship is exactly linear in $RD_{DZ0i}$. In the commonly encountered situation involving a dichotomous confounder, then for all values of $RD_{DZ0i}$ and given $P_{Z1i} \geq P_{Z0i}$, the bounding interval for $Bias_{RD}$ (in its entire nonpositive and nonnegative range) becomes:

$$(P_{Z02} - P_{Z12}) \leq Bias_{RD} \leq (P_{Z12} - P_{Z02}) \quad [7]$$

or equivalently $(P_{Z11} - P_{Z01}) \leq Bias_{RD} \leq (P_{Z01} - P_{Z11})$.

In the foregoing example, the bound width is 0.15 in either direction, attained when every $RD_{DZ0i}$ reaches its extreme value of 1 (or −1). To use these bounds, however, the investigator must have the background knowledge or be ready to assume that, had the confounder-outcome RD been available, it would have been in the direction chosen by the investigator. This is a major weakness of using bounds. That said, compared with doing nothing, a bounding analysis might be a useful back-of-the-envelope method for gauging the potential magnitude of bias from uncontrolled confounding in the primary analysis.

**Unknown confounder prevalence among the exposed ($P_{Z1i}$):** Sometimes, the prevalence of the confounder among the exposed cannot be obtained from external studies. It is still possible to calculate bounds for the unknown bias factor either over the entire range of the unknown parameter $P_{Z1i}$ or with restrictions such as assuming that $P_{Z1i} \geq P_{Z0i}$ (or $P_{Z1i} \leq P_{Z0i}$). Given only $RD_{DZ0i}$ and $P_{Z0i}$, we can define the 'maximum' bounds on $Bias_{RD}$ by using limit analysis or simply observing that the next most informative expression for the bias formula will be the one that closely approximates the bias expression in form, that is, *sans* the unknown parameter (see Appendix): If $RD_{DZ0i} \geq 0$ and assuming $P_{Z1i} \geq P_{Z0i}$ (or equivalently $RD_{DZ0i} \leq 0$ and assuming $P_{Z1i} \leq P_{Z0i}$), then

$$0 \leq Bias_{RD} \leq \max_{j}\{RD_{DZ0j} - \sum_{i=1}^{K} RD_{DZ0i}P_{Z0i}\} \quad [8]$$

or if $RD_{DZ0i} \leq 0$ and assuming $P_{Z1i} \geq P_{Z0i}$ (or equivalently $RD_{DZ0i} \geq 0$ and assuming $P_{Z1i} \leq P_{Z0i}$), then

$$\min_{j}\{RD_{DZ0j} - \sum_{i=1}^{K} RD_{DZ0i}P_{Z0i}\} \leq Bias_{RD} \leq 0. \quad [9]$$

Figure 2 is a graphical illustration of the nonnegative $Bias_{RD}$ together with the bound given by expression [8]. Along the plotted path, the unknown true bias rises to meet that bound exactly at the upper end of the range of $P_{Z13}$. As in Figure 3, that endpoint lies outside the region where $P_{Z1i} \geq P_{Z0i}$ holds in every stratum, since it sets $P_{Z12} = 0$ while $P_{Z02} = 0.10$; the largest bias compatible with the stated ordering is 0.105, with $P_{Z13} = 0.90$ and $P_{Z12} = 0.10$. The bound, therefore, remains valid but is approached rather than attained once the ordering is imposed.

**Unknown confounder prevalence among the unexposed ($P_{Z0i}$):** The confounder prevalences ($P_{Z0i}$) among those unexposed to the exposure X = 1 are probably more likely to be externally available than those among the exposed ($P_{Z1i}$). When this is not the case and only $RD_{DZ0i}$ and $P_{Z1i}$ are available, then, following our previous reasoning, the appropriate 'maximum' bounds for $Bias_{RD}$ are definable (see Appendix). If $RD_{DZ0i} \geq 0$ and assuming $P_{Z1i} \geq P_{Z0i}$ (or equivalently $RD_{DZ0i} \leq 0$ and assuming $P_{Z1i} \leq P_{Z0i}$), then

$$0 \leq Bias_{RD} \leq \max_{j}\{\sum_{i=1}^{K} RD_{DZ0i}P_{Z1i} - RD_{DZ0j}\} \qquad [10]$$

or if $RD_{DZ0i} \leq 0$ and assuming $P_{Z1i} \geq P_{Z0i}$ (or equivalently $RD_{DZ0i} \geq 0$ and assuming $P_{Z1i} \leq P_{Z0i}$), then

$$\min_{j}\{\sum_{i=1}^{K} RD_{DZ0i}P_{Z1i} - RD_{DZ0j}\} \leq Bias_{RD} \leq 0. \qquad [11]$$

Figure 3 is a graph of the bound given by expression [10], where $RD_{DZ0i} \leq 0$ and $P_{Z1i} \leq P_{Z0i}$ is assumed. For illustration, $RD_{DZ03} = -0.15$, $RD_{DZ02} = -0.12$, $P_{Z13} = 0.10$, $P_{Z12} = 0.10$, $P_{Z11} = 0.80$, and $P_{Z03}$ was varied from 0.10 to 1.00 with $P_{Z02}$ from 0.10 to 0. The bound takes the value 0.123. It is not attained inside the region where $P_{Z1i} \leq P_{Z0i}$ holds at every stratum, because the vertex $P_{Z03} \to 1$ forces $P_{Z02} = P_{Z01} = 0$; the largest bias compatible with the stated ordering is 0.12, at $P_{Z03} = 0.90$ and $P_{Z02} = 0.10$.

**DISCUSSION**

Earlier literature has shown that risk ratio bias expressions appear to plateau, bounded by certain combinations of their constituent parameters (4, 7, 21). The risk difference bias expression behaves differently: [1] is linear in each of its constituent parameters, so $Bias_{RD}$ does not plateau, and its extremes are attained only at the boundaries of the unknown parameter's range. We have derived upper and lower bounds for the risk-difference bias in cohort studies when one of the three parameters required to calculate the bias factor is unknown. These bounds represent the 'maximum' (or 'minimum') limits for the bias whenever one (unknown) parameter is allowed to vary over its range. The bias will never exceed these limits, regardless of how extreme the varying parameter becomes.

The bounds derived here cannot be narrowed without additional information, because each is attained at a vertex of the simplex of the unknown parameter. Where an ordering such as $P_{Z1i} \leq P_{Z0i}$ is additionally assumed, that vertex can fall outside the assumed region, so the bound stays valid but need not be attained, as Figures 2 and 3 illustrate. We caution against narrowing them by substituting a stratum-wise maximum, such as $max_i[P_{Z1i} - P_{Z0i}]$ or $max_i[RD_{DZ0i}P_{Z0i}]$, for the corresponding sum. For ratio measures, such a substitution is licensed because a ratio of weighted averages is bounded by the largest of the component ratios; $Bias_{\mathrm{RD}}$, however, is a sum over confounder strata, and once the confounder has more than two categories, the stratum-wise maximum can fall well below the attainable bias. It is, in any case, easy to cheat by choosing the narrowest candidate if this preserves the investigator's results and by rejecting the wider 'maximum' bound if it explains the results away. The well-informed reader who knows how to estimate these bounds could, nonetheless, estimate the 'maximum' bound and refrain from betting on the investigator's optimistic results until more evidence accrues to enable the reader to update their beliefs about the putative association or causal effect.

The bounds presented here do not necessarily account for the effects of noncompliance, nonrandom or selective exposure, and other biases on the risk difference, insofar as such biases do not result in imbalances in confounder prevalences between the exposed and unexposed, or in the confounder-outcome risk difference. Neither these bounds nor previously published ones should be construed as alternatives to a thorough sensitivity analysis conducted using Monte

Carlo or Bayesian risk assessment techniques.[22-25] We recommend restricting bounding analysis to back-of-the-envelope calculations to obtain initial insights and not using it as a shortcut to full probabilistic analysis, which would yield thorough assessments of uncontrolled confounding bias, random error, and other associated uncertainties. Bounding analysis is not a panacea for obtaining reliable information on all the required parameters to enable a fuller bias analysis. That said, bounds can come in handy for early diagnosis of potential bias in population and demographic analyses and can prompt investigators to delve deeper to assess the biases. More work is needed to assess how well these bounds perform in real-world health sciences research settings.

**Acknowledgment**

This work was initially supported by a Rubicon fellowship (grant 825.06.026) from the Board of the Council for Earth and Life Sciences (ALW) of the Netherlands Organisation for Scientific Research (NWO). This work also benefited from a seed grant from the California Center for Population Research (CCPR), which is supported by CCPR's Population Research Infrastructure Grant P2C from NICHD, P2C-HD041022. Research reported in this publication was supported by the National Center for Advancing Translational Science (NCATS) of the National Institutes of Health under the UCLA Clinical and Translational Science Institute grant number UL1TR001881. The views and findings are entirely the author's and should not be ascribed to the governments of the Netherlands or the United States.

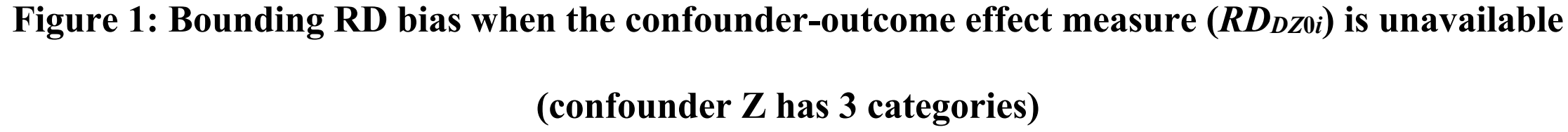

**Figure 1: Bounding RD bias when the confounder-outcome effect measure ($RD_{DZ0i}$) is unavailable (confounder Z has 3 categories)**

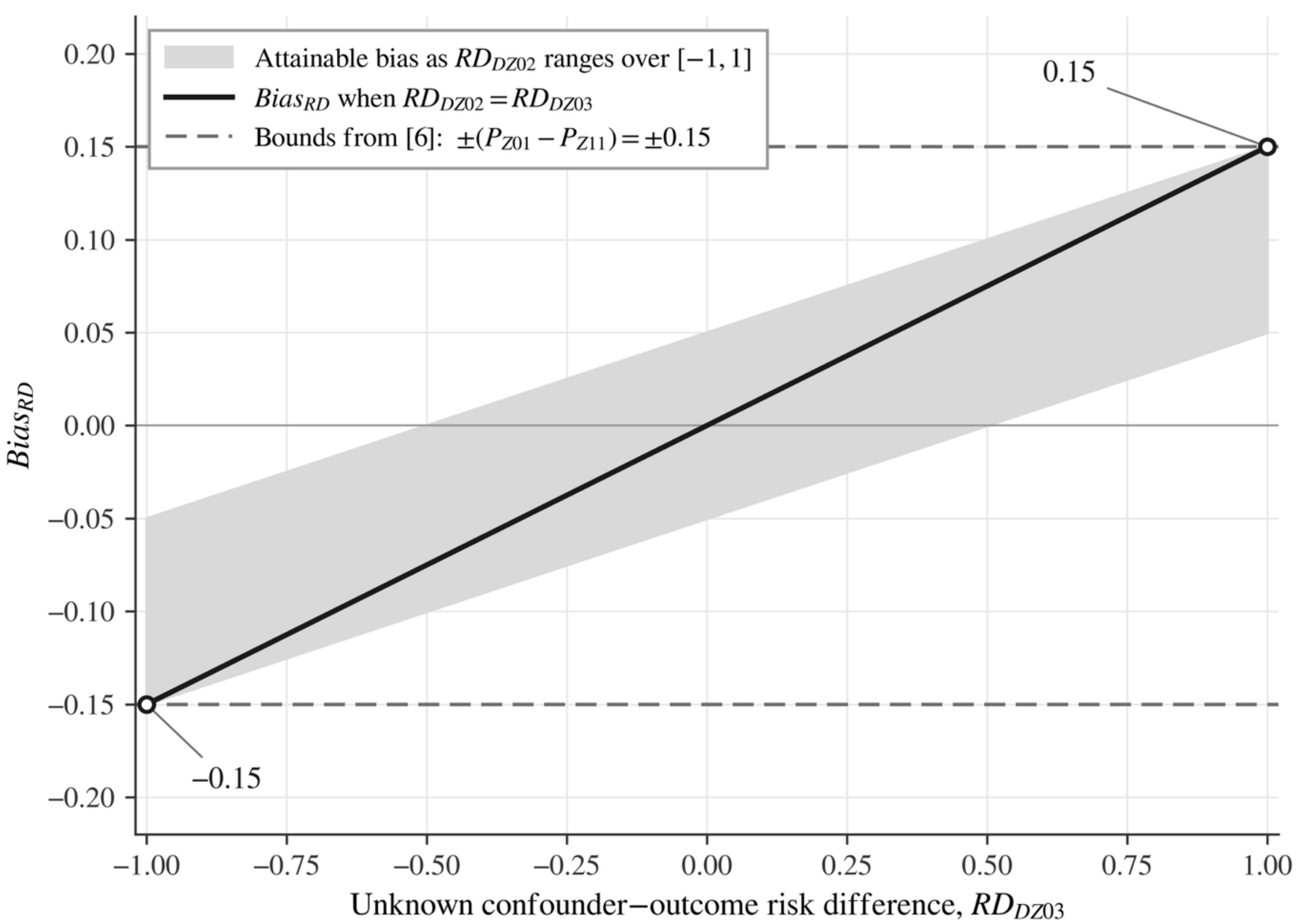

Assumed bias parameters for figure 1: $P_{Z13} = 0.50$, $P_{Z12} = 0.40$, $P_{Z11} = 0.10$, $P_{Z03} = 0.40$, $P_{Z02} = 0.35$, $P_{Z01} = 0.25$. The figure displays $Bias_{RD}$ as $RD_{DZ03}$ varies along the horizontal axis, with the shaded region representing all attainable bias values as $RD_{DZ02}$ ranges over [−1, 1]. The solid line corresponds to $RD_{DZ02} = RD_{DZ03}$. The bounds of expression 6, $\pm(P_{Z01} - P_{Z11}) = \pm0.15$, are attained at the two corners. Since $Bias_{RD}$ is linear in each $RD_{DZ0i}$, the relationship is linear and symmetric around zero.

**Figure 2: Bounding RD bias when only the confounder prevalence among the exposed ($P_{Z1i}$) is unavailable**

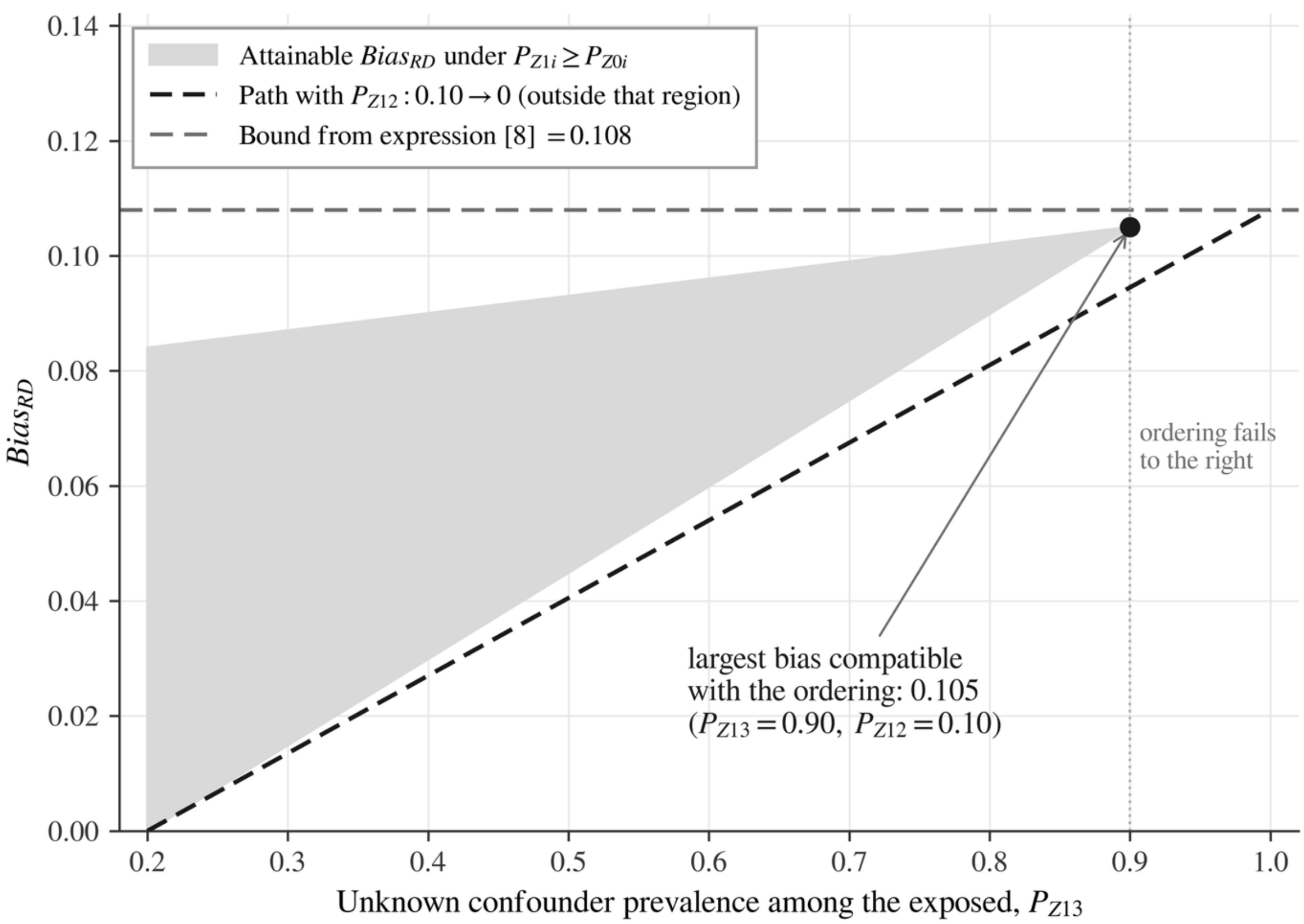

Assumed bias parameters for figure 2: $RD_{DZ03} = 0.15$, $RD_{DZ02} = 0.12$, $P_{Z03} = 0.20$, $P_{Z02} = 0.10$, and $P_{Z01} = 0.70$. The shaded region is the set of attainable $Bias_{RD}$ values as $P_{Z12}$ ranges over its feasible values, while $P_{Z1i} \geq P_{Z0i}$ holds in every stratum, which requires $P_{Z13} \leq 0.90$. The dashed line is the path along which $P_{Z12}$ falls from 0.10 to 0; it leaves that region and reaches the bound from expression [8], 0.108, only at $P_{Z13} = 1$. Within the region, the largest attainable bias is 0.105, with $P_{Z13} = 0.90$ and $P_{Z12} = 0.10$.

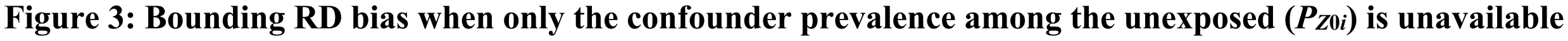

**Figure 3: Bounding RD bias when only the confounder prevalence among the unexposed ($P_{Z0i}$) is unavailable**

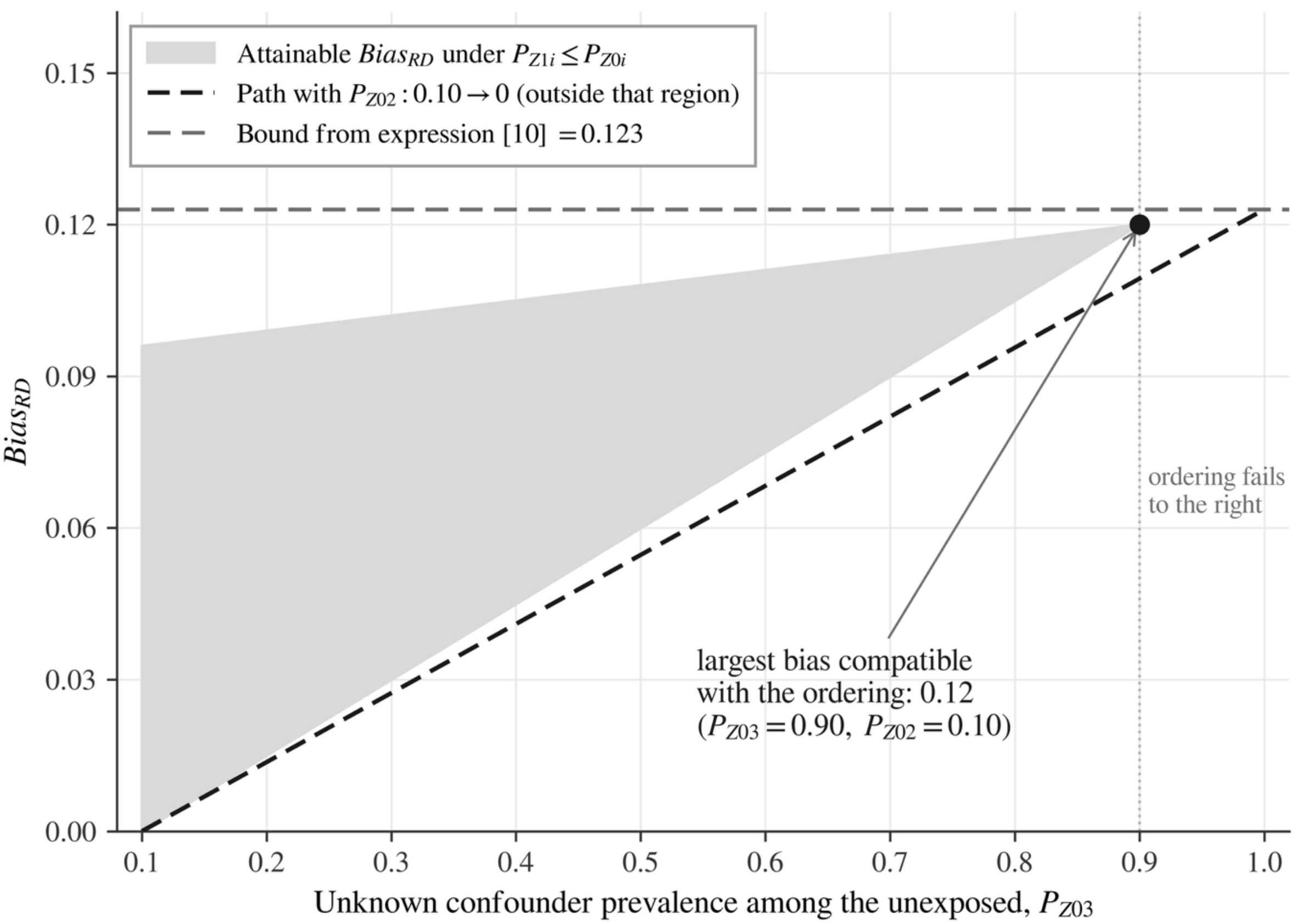

Assumed bias parameters for figure 3: $RD_{DZ0i} \leq 0$ and $P_{Z1i} \leq P_{Z0i}$; $RD_{DZ03} = -0.15$, $RD_{DZ02} = -0.12$, $P_{Z13} = 0.10$, $P_{Z12} = 0.10$, $P_{Z11} = 0.80$. The shaded region is the set of attainable bias values as $P_{Z02}$ ranges over its feasible values, while $P_{Z1i} \leq P_{Z0i}$ holds in every stratum, which requires $P_{Z03} \leq 0.90$. The dashed line is the path along which $P_{Z02}$ falls from 0.10 to 0; it leaves that region and reaches the bound from expression [10], 0.123, only at $P_{Z03} = 1$. Inside the region, the largest attainable bias is 0.12, at $P_{Z03} = 0.90$ and $P_{Z02} = 0.10$.

## APPENDIX

Suppose $RD_{DZ0i} > 0$ is unknown. Let each non-reference level $RD_{DZ0i}$ approach its maximum value. With level $i = 1$ as the reference category and $RD_{DZ01} = 0$, then for all $P_{Z1i} \geq P_{Z0i}$, i = 2, …, $K$,

$$\lim_{RD_{DZ0i} \to 1, i = 2,\ldots,K} \sum_{i=1}^{K} RD_{DZ0i}(P_{Z1i} - P_{Z0i})$$

$$= [0 \cdot (P_{Z11} - P_{Z01})] + [1 \cdot (P_{Z12} - P_{Z02})] + \ldots + [1 \cdot (P_{Z1K} - P_{Z0K})]$$

$$= \sum_{i=2}^{K} (P_{Z1i} - P_{Z0i}). \qquad [A1]$$

Therefore, for nonnegative bias (where $P_{Z1i} \geq P_{Z0i}$ and assuming $RD_{DZ0i} \geq 0$), $0 \leq Bias_{RD} \leq \sum_{i=2}^{K} (P_{Z1i} - P_{Z0i})$.

The inequality or signs can be manipulated to account for different scenarios of the relationship between the obtained $P_{Z1i}$ and $P_{Z0i}$, and for background knowledge about the direction of $RD_{DZ0i}$. See main text.

If $P_{Z1i}$ were unavailable, the investigator could still put bounds on the bias factor. Let $P_{Z1j} \to 1$ for one category $j$, and then

$$\lim_{P_{Z1j} \to 1} \sum_{i=1}^{K} RD_{DZ0i}(P_{Z1i} - P_{Z0i}) = \lim_{P_{Z1j} \to 1} \sum_{i=1}^{K} RD_{DZ0i}P_{Z1i} - \sum_{i=1}^{K} RD_{DZ0i}P_{Z0i}$$

$$= [RD_{DZ01} \cdot 0 + \ldots + RD_{DZ0j} \cdot 1 + \ldots + RD_{DZ0K} \cdot 0] - \sum_{i=1}^{K} RD_{DZ0i}P_{Z0i}$$

$$= RD_{DZ0j} - \sum_{i=1}^{K} RD_{DZ0i}P_{Z0i}$$

$$\leq \max_{j}\{RD_{DZ0j} - \sum_{i=1}^{K} RD_{DZ0i}P_{Z0i}\}. \qquad [A2]$$

The last line, $\max_{j}\{RD_{DZ0j} - \sum_{i=1}^{K} RD_{DZ0i}P_{Z0i}\}$, recognizes that the $j$th category of the uncontrolled confounder (or any similar category in which $RD_{DZ0i}$ would have taken its most extreme value) will yield the maximum value of this expression. This is the maximum upper bound for nonnegative bias. Estimating the reverse inequalities for nonpositive bias is straightforward.

Furthermore, let the unknown $P_{Z0j} \to 1$, and it can be shown that the maximum upper bound of the bias expression, when only $RD_{DZ0i}$ and $P_{Z1i}$ are known, is similarly given by

$$\lim_{P_{Z0j} \to 1} \sum_{i=1}^{K} RD_{DZ0i}(P_{Z1i} - P_{Z0i}) = \sum_{i=1}^{K} RD_{DZ0i}P_{Z1i} - RD_{DZ0j}$$

$$\leq \max_{j}\{\sum_{i=1}^{K} RD_{DZ0i}P_{Z1i} - RD_{DZ0j}\}. \quad \text{[A3]}$$